\documentclass{ptephy}
\preprintnumber{XXXX-XXXX}

\usepackage{boites}
\numberwithin{equation}{section}
\usepackage{amsmath}	% required for `\align' (yatex added)
\begin{document}

\title{Gauge Symmetry in the Large-amplitude Collective Motion of Superfluid Nuclei}

\author{\name{\fname{Koichi} \surname{Sato}}{1} 
%\name{\fname{Second} \surname{Author}}{2,\dag}, 
%and \name{\fname{Third} \surname{Author}}{3,\ast,}
%\thanks{These authors contributed equally to this work}
}

\address{\affil{1}{RIKEN Nishina Center, Wako 351-0198, Japan}
%\affil{2}{Second author address}
%\affil{3}{Third author address}
\email{satok@ribf.riken.jp}}

\begin{abstract}%
The adiabatic self-consistent collective coordinate (ASCC) method is a
 practical method for the description of large-amplitude collective
 motion in atomic nuclei with superfluidity and
an advanced version of the adiabatic time-dependent Hartree-Fock-Bogoliubov theory.
We investigate the gauge symmetry in the ASCC method
on the basis of the theory of constrained systems. 
The gauge symmetry in the ASCC method is originated from 
the constraint on the particle number in the collective Hamiltonian, 
and it is partially broken by the adiabatic expansion.
The validity of the adiabatic expansion under the general gauge transformation is also discussed.
\end{abstract}

\subjectindex{xxxx, xxx}

\maketitle

%\tableofcontents

\section{Introduction}\label{sec1}
 It has been one of the long-standing open problems in nuclear physics
to construct a microscopic theory of 
large-amplitude collective motion of atomic nuclei~\cite{Matsuyanagi2010, Klein1991}.
One way to describe the large-amplitude collective motion is 
the adiabatic time-dependent Hartree-Fock(-Bogoliubov) (ATDHF(B)) theory. 
Several versions of the ATDHF have been proposed so far 
but they have encountered 
various difficulties such as non-uniqueness of the solution 
(See \cite{Klein1991} for a review).

The adiabatic self-consistent collective coordinate (ASCC) method
~\cite{Matsuo2000}
is an adiabatic approximation to the self-consistent
collective coordinate (SCC) method
and can be regarded as an advanced version of the ATDHFB theory.
The SCC method was originally formulated by Marumori et al.\cite{Marumori1980}
on the basis of the TDHF theory 
and then was extended to include the pairing correlation~\cite{Matsuo1986a}.
It enables us to extract a small-dimensional collective submanifold 
embedded in the huge-dimensional phase space of the TDHF(B) theory.
The solution of the SCC basic equations first proposed 
was the perturbative expansion with respect to the 
collective coordinates and momenta around the HFB states.
Thus it was difficult to apply it to the large-amplitude collective motions.

The ASCC method is a non-perturbative scheme of solving the basic
equations of the SCC method  based on the adiabatic expansion 
with respect to the collective momenta.
By taking up to the second-order terms of the collective momenta,
the ASCC method overcomes the difficulty the ATDHF theory of Villars
encountered in obtaining a unique solution~\cite{Villars1977}.
The aim of the ASCC method is to extract small-dimensional collective
degrees of freedom the system itself chooses from the large-dimensional
TDHFB phase space and construct the dynamics on the collective submanifold,
and it is more ambitious than that of Baranger-V{\'e}n{\'e}roni's ATDHF theory~\cite{Baranger1978}.

The one-dimensional version of the ASCC method has been successfully applied
to the shape coexistence/fluctuation phenomena in proton-rich Se and Kr 
isotopes~\cite{Hinohara2008, Hinohara2009}. 
(Here, we mean by the $D$-dimensional ASCC method that
the dimension of the collective coordinate $q$ is $D$.)
The approximate version of the two-dimensional ASCC method is called 
the constrained Hartree-Fock-Bogoliubov plus
local quasiparticle random phase approximation 
(CHFB+LQRPA) method~\cite{Hinohara2010b}
and is employed to microscopically derive 
the five-dimensional quadrupole collective Hamiltonian.
The CHFB+LQRPA method has been also applied 
to large-amplitude quadrupole collective dynamics
in $^{68-72}$Se~\cite{Hinohara2010b},  $^{72-76}$Kr~\cite{Sato2011}, 
$^{110}$Mo~\cite{Watanabe2011}, $^{26}$Mg~\cite{Hinohara2011a},
$^{30-34}$Mg~\cite{Hinohara2011},
$^{130-134}$Ba and $^{128-132}$Xe~\cite{Hinohara2011},
$^{58-68}$Cr~\cite{Yoshida2011} and 
$^{58-66}$Cr~\cite{Sato2012}.

In solving the basic equations of the ASCC method, 
%(moving-frame HFB and moving-frame QRPA equations), 
Hinohara et al.~\cite{Hinohara2007} encountered a numerical instability 
%in solving the moving-frame HFB \& QRPA equations 
and found that it was caused by some continuous symmetry of 
the basic equations.
More precisely, they found that
the basic equations of the ASCC method, i. e.,
the canonical-variable conditions, the moving-frame HFB \& QRPA
equations are invariant under the transformation,
\begin{align}
 \hat Q &\rightarrow \hat Q +\alpha \tilde N , \\
 \hat \Theta &\rightarrow \hat \Theta + \alpha \hat P, \\
 \lambda &\rightarrow  \lambda -\alpha \partial_q V,\\
 \partial_q\lambda &\rightarrow  \partial_q\lambda -\alpha C .
\end{align}
This transformation involves the gauge angle $\varphi$ and changes the phase of the state vector. 
In this sense, Hinohara et al. called it the ''gauge'' symmetry.
They proposed a gauge-fixing prescription 
to remove the redundancy associated with the gauge symmetry 
and successfully applied it to the multi-$O(4)$
model \cite{Hinohara2007} and the shape coexistence in proton-rich Se and Kr 
isotopes with the pairing-plus-quadrupole model
~\cite{Hinohara2008, Hinohara2009}.

In this paper, we investigate this symmetry on the basis of
the theory of constrained systems, which was 
initiated by Dirac and Bergmann \cite{Dirac1950, Anderson1951, Dirac1964}. 
As is well known, the gauge symmetry is associated
with constraints which are originated from the singularity of the Lagrangian.
In the ASCC method formulated in Refs. \cite{Matsuo2000, Hinohara2007},
the collective Hamiltonian  is expanded up to $O(n)$.
In fact, the linear term of the particle number in the collective Hamiltonian 
can be regarded as a constraint, 
and it leads to the gauge symmetry in the ASCC method. 
%is associated with this constraint on the particle number.
Moreover, to take into account pairing rotational modes,
the gauge angle is explicitly included in the ASCC method.
As we shall see, the $\varphi$ dependence of the state vector in
Eq. (\ref{eq:hin2.3}) plays an important  role for the gauge symmetry.
%We discuss possible gauge transformations and the validity 
%of the adiabatic expansion in the ASCC method under those transformations.
As mentioned above, a large part of the discussion in this paper is
based on the Dirac-Bergmann theory for constrained systems,
on which the reader is referred to, e.g., \cite{Dirac1964, Hanson1976, Henneaux1994}.

This paper is organized as follows. 
In Sec. \ref{sec:HT}, we introduce the collective Hamiltonian and show that it can
be regarded as a Hamiltonian with a primary constraint $n=0$
and that the consistency condition for the primary constraint is fulfilled.
In Sec. 3, the possible forms of generators of the gauge transformation
and the validity of the adiabatic expansion are discussed.
A brief discussion on the gauge fixing is given in Sec. \ref{sec:fixing}.
In Sec. 5, we present the basic equations of the ASCC method,
and in Sec. 6 we illustrate some examples of simple gauge transformations 
as well as the most general one. We see that, 
while the gauge symmetry is conserved in the equation of collective submanifold,
it is partially broken by the adiabatic expansion in the moving-frame 
HFB \& QRPA equations.
In Sec. 7, it is shown that there is no gauge symmetry 
in the case where the collective Hamiltonian is expanded up to $O(n^2)$. 
Concluding remarks are given in Sec. 8.

In this paper, we consider the one-dimensional ASCC method with a single
component for simplicity.
However, the extension to the multi-dimensional and 
multi-component (e. g., protons and neutrons in nuclei) 
cases is straightforward.

%This \LaTeX\ class file is available for the authors to prepare
%the manuscript for PTEPHY Journal. It is assumed that
%the authors are familiar with either plain \TeX, \LaTeX,\
%\AmS-\TeX\ or a standard \LaTeX\ set-up, hence, only the
%essential points are described in this document. To get more
%details please go through the \textit{\LaTeX\ User's Guide} or
%\textit{The not so short introduction to \LaTeXe} (which is available online).

\section{Hamiltonian and Constraints} \label{sec:HT}
%For simplicity, we shall consider one-dimensional ASCC method.
%The extension to the multi-dimensional cases is straightforward.
We follow the formulation in Refs. \cite{Matsuo2000} and \cite{Hinohara2007}. 
We assume the following form of the state vector.
\begin{align}
 | \phi(q,p,\varphi, n) \rangle &= e^{-i \varphi \tilde N
  | }|\phi(q,p,n)\rangle, \label{eq:hin2.3}\\  
 | \phi(q,p, n) \rangle &= e^{i \hat G }|\phi(q)\rangle,  \\
G(q,p,n)&=p\hat Q(q) +n\hat \Theta (q).
\end{align}
$\varphi$ is the gauge angle conjugate to the particle number $n=N-N_0$.
We measure the particle number from a reference value $N_0$ and 
define $\tilde N=\hat N-N_0$.
The collective Hamiltonian is defined by
\begin{align}
\mathcal{H}(q,p,n) &=  \langle \phi (q,p,\varphi, n) | \hat H |\phi(q,p,\varphi,n)\rangle  \notag \\ 
&=  \langle \phi (q,p,n) | \hat H |\phi(q,p,n)\rangle  \notag \\ 
&=\langle \phi (q) |e^{-i\hat  G (q,p,n)} \hat H e^{i\hat  G (q,p,n)}|\phi(q)\rangle , 
\end{align}
and expanded as below:
\begin{align}
\mathcal{H}(q,p,n) &=  V(q)+\frac{1}{2}B(q)p^2 +\lambda n,  \\
%\end{align}
%\begin{align}
 V(q)&=\langle\phi (q)|\hat H |\phi(q)\rangle, \\
 B(q)&= %\langle\phi (q)|[\hat H, i\hat  Q^{(2)}] |\phi(q)\rangle 
      -\langle\phi (q)|[[\hat H, \hat  Q],\hat Q] |\phi(q)\rangle, \\
 \lambda(q)&=\langle\phi (q)|[\hat H,i\hat \Theta] |\phi(q)\rangle.
 \label{eq:lambda def}%\\
% D(q)&=\langle\phi (q)|[\hat H, i\hat  \Theta^{(2)}] |\phi(q)\rangle 
%      -\langle\phi (q)|[[\hat H, \hat  \Theta^{(1)}],\hat \Theta^{(1)}] |\phi(q)\rangle 
\end{align}
Note that the canonicity conditions (\ref{eq:canonicity
1})-(\ref{eq:canonicity 4}) imply that $(q,\varphi)$ should be regarded as coordinates 
and $(p,n)$ as their conjugate momenta.

%It is convenient to denote $\mathcal{H}$ as $H_T$ %define the total Hamilton
For convenience, we shall put
\begin{align}
\mathcal{H}(q,p,\varphi,n) &= \frac{1}{2}B(q)p^2 + V(q, \varphi)+\lambda
 n=H +\lambda n =H_T .\label{eq:HT}
\end{align}
We call  $H_T=\mathcal{H}$ the total Hamiltonian 
and distinguish it  from the Hamiltonian $H$.
This can be regarded as a system with the constraint $n=0$, 
and $\lambda$ is a Lagrange multiplier.
We have allowed the potential $V$ to depend on $\varphi$ above
in order to make it easy to see the number of the degrees of freedom.
However, we mainly consider the case with $V=V(q)$ below
as the collective potential in the ASCC method is independent of $\varphi$.
Hereinafter, we denote $(q,\varphi)=(q^1, q^2)$ and  $(p,n)=(p_1, p_2)$ also.
(This total Hamiltonian is of the form of a weakly reducible
system~\cite{Sugano1983}. Note that because $\lambda$ depends on $q$, 
the degree of freedom of $(\varphi, n)$ is not completely decoupled 
from that of $(q,p)$ even if $V=V(q)$.)

With this total Hamiltonian, the time evolution of a physical quantity $f(q,p,\varphi,n ,t)$ is given by
\begin{align}
\frac{df}{dt} =\left\{ f, H_T \right\} + \frac{\partial f}{\partial t}.
\end{align}
Here $\{\cdot  , \cdot\}$ denotes the Poisson bracket.
It is noteworthy that Eq. (2.9) of \cite{Matsuo2000} is the canonical equations for $V=V(q)$.

In the Lagrange formalism, the Lagrangian of this system is given by
\begin{align}
 L=\frac{1}{2B(q)}(\dot q^1)^2-V(q,\varphi) \label{eq:Lagrangian},
\end{align}
and then the rank of the Hessian $(\partial^2 L /\partial \dot q^i \partial \dot
q^j)$ is one.
Hence, it leads to one constraint
\begin{align}
p_2=\frac{\partial L}{\partial \dot q^2}=n=0.
\end{align}
Its time derivative is given by
\begin{align}
\dot n=\left\{ n, H_T \right\}=-\partial_\varphi V \label{eq:n dot}.
\end{align}
Thus, the consistency condition, i. e., 
the condition that the constraint be preserved in time, 
is automatically fulfilled 
if $V(q,\varphi)=V(q)$.

Below we consider the $V=V(q)$ case unless otherwise noted.
Then we have only one constraint $n=0$, and it is a first-class one.
From the above, one sees that our system has one gauge degree of freedom.
(In the ASCC method, the consistency condition is guaranteed by the
$\varphi$ dependence of the state vector (\ref{eq:hin2.3})).
The expansion of the collective Hamiltonian up to $O(n^2)$ is
investigated in Sec. \ref{sec:O(n^2)}.
 
\section{Generators of gauge transformations} \label{sec:generators}
We shall consider infinitesimal gauge transformations.
A generator of a gauge transformation can be always written 
as a linear combination of the constraints of the first
class~\cite{Sugano1982, Castellani1982,Sugano1991}. 
We can write the generator $G$ as
\begin{align}
 G=\epsilon(q,p,\varphi,n,t)n \label{eq:G}.
\end{align}
Here $\epsilon$ is an infinitesimal function.
%According to generalities of the gauge transformations,
%$\epsilon$ can depend on time explicitly, i.e.,  $\epsilon=\epsilon(t)$.    
%However, we assume that the state vector (\ref{eq:hin2.3}) does not
%depend on time explicitly, 
%and %if we allow $\epsilon=\epsilon(t)$
%the explicit time dependence of $\epsilon$ leads to that of the state vector. 
%Therefore, we concentrate on  $\epsilon$ which is not dependent on time
%explicitly here. 
From the stationary condition of $G$,
\begin{align}
\dot G=\left\{ G, H_T \right\} + \frac{\partial G}{\partial t} \approx 0.
%\dot G=\left\{ G, H_T \right\} \approx 0 .
\end{align}
Here, the symbol $\approx$ denotes the weak equality and the equation above means
\begin{align}
 \dot G = 0 \,\,\,\,\, ({\rm mod} \,\,\, n).
\end{align}
Because the Lagrange multiplier in $H_T$ is an arbitrary function, 
this stationary condition is divided into two parts as below:
\begin{align}
&\left\{ G, H \right\} +\frac{\partial G}{\partial t}\approx 0  \label{eq:dGdt1},\\
&\left\{ G, n \right\} \approx 0 \label{eq:dGdt2}.
\end{align}
By substituting (\ref{eq:G}) into (\ref{eq:dGdt1}), we obtain
%\begin{align}
%&\left\{\epsilon, H \right\}n + \epsilon\left\{n, H \right\} + \frac{\partial \epsilon }{\partial t}n  \approx 0.  
%\end{align}
%we obtain
\begin{align}
\left( \left\{\epsilon, H \right\} + \frac{\partial \epsilon }{\partial
 t}\right)n  \approx 0 , 
%\left\{\epsilon, H \right\} n  \approx 0 . 
\end{align}
where we have used $\partial H/\partial \varphi=0$.
Hence, if $\epsilon$ is not a singular function of $n$, 
Eq. (\ref{eq:dGdt1}) holds for arbitrary $\epsilon(q,p, \varphi, n, t)$.
Eq. (\ref{eq:dGdt2}) leads to
\begin{align}
\left\{\epsilon, n\right\}n + \epsilon\left\{n, n\right\} =\left\{\epsilon, n\right\}n \approx 0, 
\end{align}
i. e., $\partial_\varphi \epsilon \, n  \approx 0$.
One sees that Eq. (\ref{eq:dGdt2}) is also satisfied if $\partial_\varphi
\epsilon$ is a regular functions of $n$.
Therefore, we concentrate on $\epsilon$ which is regular with respect to
$n$. 

Then, the infinitesimal gauge transformation of $(q_i, p_i)$ is given by
\begin{align}
\delta q^i = \left\{q^i, G\right\}, \,\,\, \delta p_i = \left\{p_i,
 G\right\} \label{eq:delta q,p},
\end{align}
from which we obtain
\begin{align}
\delta q &=n\partial_p \epsilon \approx 0, \,\,\, \delta p
 =-n\partial_q\epsilon \approx 0\\ 
\delta \varphi &=\epsilon +n\partial_n \epsilon \approx \epsilon, \,\,\,
 \delta n =-n\partial_\varphi \epsilon \approx 0.
\end{align}
As $\epsilon$ is an arbitrary function of $(q,p,\varphi,n,t)$,
in particular, of $p$, 
%the terms of $O(p^m)\,\,(m \geq 1)$ 
the linear and higher-order terms of $p$
are mixed only into $\varphi$
by this gauge transformation.
This is important for the adiabatic expansion in the ASCC method to make sense.
(Otherwise, one needs to fix the gauge in order not to mix terms of
different orders of $p$. )

\subsection{Example 0: $\epsilon =const. $}
The simplest example of the gauge transformation is the case with $\epsilon =const.$
In this case, $G=\epsilon n$ generates the transformation
\begin{align}
\delta q=0, \,\,  \delta p=0, \,\, \delta n=0,  \,\, \delta
 \varphi=\epsilon .
\end{align}
This transforms $\varphi$ as $\varphi \rightarrow \varphi +\epsilon$,
and it corresponds to the global $U(1)$ gauge transformation
changing the phase of the state vector.
We shall see the relation with the phase transformation of the BCS-type
wave function below.

We expand the state vector
$| \phi(q,p,N) \rangle$ in terms of the particle number basis.
\begin{align}
&e^{-i\varphi \hat N}| \phi(q,p,N) \rangle
=e^{-i\varphi \hat N} \sum a_N |N \rangle =\sum e^{-i\varphi N} a_N |N \rangle \notag \\
=& \cdots + a_{N-2} e^{-i\varphi (N-2)}| N-2\rangle + a_{N} e^{-i\varphi N}| N\rangle + a_{N+2} e^{-i\varphi (N+2)}| N+2\rangle +\cdots.
\end{align}
For simplicity, we have omitted the degrees of freedom of $(q,p)$ and
the overall phase $e^{i\varphi N_0}$ here.
The BCS wave function is written as
\begin{align}
&| \Psi \rangle=\Pi_\nu \left( u_{\nu}+ |v_{\nu}|e^{i\tilde \varphi} c_{\nu}^\dagger
 c_{\nu}^\dagger \right) | -\rangle,
\end{align}
and the phase of the  component with the particle number $N$  is
$e^{i\tilde \varphi N/2}$~\cite{Tinkham2004, Brink2005}.
By equating $\tilde \varphi/2$ with $-\varphi$, 
one can see that the above gauge transformation  
corresponds to the gauge transformation of the BCS wave function 
$\tilde \varphi \rightarrow \tilde \varphi + \alpha$.

\section{Gauge fixing} \label{sec:fixing}
We shall briefly discuss the gauge fixing
with a gauge fixing condition $\chi \approx 0$.
To fix the gauge, $\chi$ should fulfill
\begin{align}
\{\chi, n \} \neq 0 .
\end{align}
The stationary condition for the gauge fixing condition $\chi$ gives
\begin{align}
\{\chi , H\}+ \lambda \{\chi, n \} \approx 0,
\end{align}
from which  $\lambda$ is determined as
\begin{align}
 \lambda = -\frac{\{\chi,H\}}{\{\chi,n\}}.\label{eq:lambda_GF}
\end{align}
If we assume that $\chi$ is independent of $(q,p)$, i. e., $\chi=\chi(\varphi, n)$,
Eq. (\ref{eq:lambda_GF}) gives $\lambda=0$. 
As an example of $\chi$ depending on $(q,p)$, we shall consider a gauge
fixing condition 
of the following form:
\begin{align}
\chi= e^\varphi e^{-\tilde \chi(q,p)}-\chi_0 .
\end{align}
%with a constant $\chi_0$.
Here we have introduced a constant $\chi_0$ to prevent $\tilde \chi$
from diverging when the condition $\chi \approx 0$ is fulfilled.
Then $\lambda$ is given by 
\begin{align}
\lambda =\{\tilde \chi, H\}, \label{eq:lamdba GF}
\end{align}
for a given $\tilde \chi$. 
%$\lambda$ is determined by Eq. (\ref{eq:lambda GF} ).
Conversely, once $\lambda$ is given, $\tilde \chi$ is determined by
solving the linear partial differential equation (\ref{eq:lamdba GF}).
For example, in the case of $\lambda = \lambda_0=const.$, 
$\tilde \chi$ is determined as $\tilde \chi=\lambda_0t/2$.
$\lambda=const.$ means $\partial_q \lambda =0$, and 
it corresponds to the QRPA gauge \cite{Hinohara2007}.
(The QRPA gauge fails at a point where $C=0$ as discussed in
Ref. \cite{Hinohara2007}. 
This situation occurs when there are more than one local minimum in the collective potential.
Let $q_1$ and $q_2$ be the positions of two minima of the potential,
where $\partial_q V=0$.
%, and denote $\lambda(q_1):=\lambda_1$ and $\lambda(q_2):=\lambda_2$. 
As $\lambda(q_1) \neq \lambda(q_2)$ in general,  the failure of the 
QRPA gauge with $\lambda=const$ is unavoidable.)

\section{Basic equations of the ASCC method}
Before moving to the discussion on the gauge symmetry 
in the ASCC method, 
we present the basic equations of the ASCC method.
For the details of the derivation  of the basic equations,
see Ref. \cite{Matsuo2000}.

\noindent
\underline{The zeroth and first order canonical-variable conditions:}
\begin{align}
\langle \phi(q)|\,\hat Q   \,|\phi(q)\rangle =0. \\
\langle \phi(q)|\,\hat P   \,|\phi(q)\rangle =0. \\
\langle \phi(q)|\,\hat \Theta   \,|\phi(q)\rangle =0. \\
\langle \phi(q)|\,\tilde N   \,|\phi(q)\rangle =0. \label{eq:O(1) can.var.cond 4}
\end{align}
\begin{align}
% O(p)
&\langle \phi(q)|\,[\hat Q, \hat P] \,|\phi(q)\rangle =i. \\
%&\langle \phi(q)|\hat     Q^{(2)}|\phi(q)\rangle =0. \\
&\langle \phi(q)|\,[\hat Q, \hat \Theta] \,|\phi(q)\rangle
 =0. \\
&\langle \phi(q)|\,[\hat Q, \tilde N] \,|\phi(q)\rangle =0. \\
% O(n)
&\langle \phi(q)|\,[\hat P, \hat \Theta] \,|\phi(q)\rangle
 =0. \label{eq:O(n) can.var.cond 1} \\
&\langle \phi(q)|\,[\hat P, \tilde N] \,|\phi(q)\rangle =0. \label{eq:O(n) can.var.cond 2} \\
&\langle \phi(q)|\,[\hat \Theta, \tilde N] \,|\phi(q)\rangle =i. \label{eq:O(n) can.var.cond 3} 
%&\langle \phi(q)|\hat     \Theta^{(2)}|\phi(q)\rangle =0.
\end{align}

\noindent
\underline{Moving-frame HFB equation:}
\begin{equation}
 \delta \langle \phi(q)|\hat H -\lambda \tilde N -\partial_q V\hat Q
  |\phi(q)\rangle =0. \label{eq:moving-frame HFB}
\end{equation}
\noindent
\underline{Moving-frame QRPA equations:}
\begin{equation}
 \delta \langle \phi(q)|[\hat H-\lambda \tilde N -\partial_q V \hat Q, i\hat Q] -B(q)\hat P 
%-\frac{1}{i}\partial_q V \hat Q^{(2)} 
|\phi(q)\rangle =0.\label{eq:moving-frame QRPA1}
\end{equation}
\begin{align}
 \delta \langle \phi(q)|
[\hat H -\lambda \tilde N -\partial_q V \hat Q, \hat P]
-iC(q)\hat Q-i\partial_q \lambda \tilde N \notag\\
-\frac{1}{2B}\left\{[[\hat H-\lambda \tilde N-\partial_q V \hat Q, \partial_q V \hat
 Q],i\hat Q] \right.
%-i \partial_q V [\hat H-\lambda \tilde N, \hat Q^{(2)}] \right.\notag \\
%\left. -\frac{i}{2}(\partial_q V)^2 [\hat Q^{(1)}, \hat Q^{(2)}]\right\}
|\phi(q)\rangle =0. \label{eq:moving-frame QRPA2}
\end{align}

The above equations are derived from the more fundamental equations by
the adiabatic expansion.
The moving-frame HFB \& QRPA equations are derived from the following equation:

\noindent
\underline{Eq. of collective submanifold:}
\begin{align}
& \delta \langle \phi(q,p,\varphi, N)|\hat H -
  \frac{\partial \mathcal{H}}{\partial p}\mathring{P}
- \frac{\partial \mathcal{H}}{\partial q}\mathring{Q}
- \frac{\partial \mathcal{H}}{\partial \varphi}\mathring{\Theta}
- \frac{\partial \mathcal{H}}{\partial N}\mathring{N}
|\phi(q,p,\varphi, N) \rangle  &=0, \label{eq:coll.man1} \\
\Longleftrightarrow & \delta \langle \phi(q,p,N)|\hat H -
  \frac{\partial \mathcal{H}}{\partial p}\mathring{P}
- \frac{\partial \mathcal{H}}{\partial q}\mathring{Q}
- \frac{\partial \mathcal{H}}{\partial N}\mathring{N}
|\phi(q,p,N) \rangle  &=0, \label{eq:coll.man2} \\
\Longleftrightarrow & \delta \langle \phi(q)|e^{-i\hat G(q)}\hat H
 e^{i\hat G(q)}-
  \frac{\partial\mathcal{H}}{\partial p}\mathring{P^\prime}
- \frac{\partial\mathcal{H}}{\partial q}\mathring{Q^\prime}
- \frac{\partial\mathcal{H}}{\partial N}\mathring{N^\prime}
|\phi(q) \rangle  &=0 ,\label{eq:coll.man3} \\
\Longleftrightarrow & \delta \langle \phi(q)|e^{-i\hat G(q)}\hat H
 e^{i\hat G(q)}
-B(q)p\mathring{P^\prime}  \notag\\
&- (\frac{1}{2}\partial_qB(q)+\partial_q V(q)+\partial_q \lambda N)\mathring{Q^\prime}
- \lambda(q)\mathring{N^\prime}|\phi(q) \rangle & =0, \label{eq:coll.man4} 
\end{align}
with 
$\mathring{P^\prime}=e^{-i\hat G}\mathring{P}e^{i\hat G},
 \mathring{Q^\prime}=e^{-i\hat Q}\mathring{Q}e^{i\hat G},$
and 
$\mathring{N^\prime}=e^{-i\hat G}\hat{N}e^{i\hat G}$.
For the first equality, $\partial \mathcal{H}/\partial \varphi=0$ is
used.

The canonical-variable conditions are derived from the canonicity conditions.

\noindent
\underline{Canonicity conditions:}
\begin{align}
   \langle \phi(q,p,\varphi,N)|\mathring{P} |\phi(q,p,\varphi, N) \rangle&
 =p+\frac{\partial s }{\partial q}, \label{eq:canonicity 1}\\
   \langle \phi(q,p,\varphi,N)| \mathring{Q}|\phi(q,p,\varphi, N) \rangle&
 =-\frac{\partial s }{\partial p}, \\
   \langle \phi(q,p,\varphi,N)|\mathring{N } |\phi(q,p,\varphi, N) \rangle&
 =N+\frac{\partial s }{\partial \varphi}, \\
         \langle \phi(q,p,\varphi,N)| \mathring{\Theta}|\phi(q,p,\varphi, N) \rangle&
 =-\frac{\partial s }{\partial N}. \label{eq:canonicity 4} 
\end{align}
Here $s$ is an arbitrary function and is set to $s=0$ in the ASCC method.
This arbitrary function $s$ is related with the generating function of a
canonical transformation as shown below~\cite{Yamamura1987}.

Let $(q^i,p_i)$ and  $(Q^{i},P_i)$ be two sets of canonical variables 
which satisfy the canonicity conditions with the arbitrary functions $s$
and $S$, respectively. Their Liouville 1-forms are connected with each
other by
\begin{align}
 p_idq^i+ds=P_idQ^i+dS.
\end{align}
This implies that $W=S-s$ is the generating function of a time-independent canonical transformation.

It is worthwhile to see its relation with a time-independent gauge
transformation. 
Assume that the canonical variables $(q^i,p_i)$ defined with $s=0$ are transformed
to $(Q^i,P_i)$ by an infinitesimal gauge transformation.
From $s=0$, it follows that $S=W$ is the generating function.
For the infinitesimal gauge transformation, we can set 
\begin{align}
 Q^i&=q^i+\delta q^i, \\
 P_i&=p_i+\delta p_i,
\end{align}
with the infinisimals $(\delta q^i, \delta p_i)$, and then it leads to
\begin{align}
 P_idQ^i=(p_i+\delta p_i)d(q^i+\delta q^i )=p_idq^i+ \delta p_idq^i+p_i
 d\delta q^i . \label{eq:PdQ}
\end{align}
Here we have omitted the second order infinitesimals.
By substituting Eq. (\ref{eq:PdQ}) into 
\begin{align}
p_idq^i&=P_idQ^i+dW(q^i,Q^i), 
\end{align}
we obtain
\begin{align}
0&=\delta p_idq^i+p_i d\delta q^i +\frac{\partial W}{\partial q^i}dq^i
 +\frac{\partial W} {\partial \delta q^i } d\delta q^i, \label{eq:delta p dq}
\end{align}
from which we read
\begin{align}
 p_i&=-\frac{\partial W} {\partial \delta q^i },\\ 
 \delta p_i&=-\frac{\partial W}{\partial q^i}.
\end{align}
By substituting  $W:=-p_i\delta q^i +W^\prime$ into Eq. (\ref{eq:delta p
dq}),
we obtain
%\begin{align}
%0=\delta p_idq^i-\delta q^i dp_i+ dW^\prime 
%\end{align}
%$i.e.,$
\begin{align}
 \delta q^i&= \frac{\partial W^\prime} {\partial p_i }=\{q^i, W^\prime\},\\ 
 \delta p_i&=-\frac{\partial W^\prime}{\partial q^i}=\{p_i, W^\prime\}.
\end{align}
From Eq. (\ref{eq:delta q,p}), one sees that $W^\prime=G$ up to a
constant.
It follows that
\begin{align}
S=W=-p_i\delta q^i +G= -p_i \{q^i, G\}+G=-p_i \partial_{p_i}G+G .
\end{align}

\section{Examples of gauge transformations} \label{sec:examples}

We shall relate the gauge transformations of $(q^i,p_i)$ with the
transformations of the operators 
$(\hat Q, \hat P, \hat \Theta ,\tilde N)$.
In Ref. \cite{Hinohara2007}, it is pointed out that,
if $[\hat N, \hat Q]=0$, the effect of the transformation of operators
\begin{align} 
\hat Q      & \rightarrow  \hat Q   +\alpha \tilde N  \label{eq:hin3.5a},\\
\hat \Theta & \rightarrow \hat \Theta + \alpha \hat P \label{eq:hin3.5b},
\end{align}
on the state vector is equivalent to that of the transformation 
%of the arguments  
\begin{align} 
q      & \rightarrow  q -\alpha n,  \label{eq:hin_gauge_q}\\
\varphi & \rightarrow \varphi - \alpha p \label{eq:hin_gauge_phi}.
\end{align}
Note that the signs of $\alpha$ are opposite.
This can be observed by applying (\ref{eq:hin3.5a})-(\ref{eq:hin3.5b}) to the state vector. 
\begin{align}
 |\phi(q,p,\varphi,n) \rangle &\rightarrow e^{-i\varphi\tilde N}e^{ip(\hat Q(q) +\alpha\tilde N) +in(\Theta(q)+\alpha \hat P(q) )}  |\phi(q)\rangle \notag \\
%&=e^{-i\varphi \tilde N}e^{ip(\hat Q(q) +\alpha\tilde N)}e^{in(\hat \Theta(q)+\alpha \hat P(q) )}  |\phi(q)\rangle +O(pn) \notag \\
%&=e^{-i\varphi \tilde N}e^{ip\alpha\tilde N}e^{ip \hat Q(q)} e^{in\hat \Theta(q)}e^{in\alpha \hat P(q) }  |\phi(q)\rangle +O(pn, n^2)\notag \\
%&=e^{-i(\varphi      -\alpha p)\tilde N}e^{ip \hat Q(q)} e^{in\hat \Theta(q)} |\phi(q-\alpha n)\rangle +O(pn, n^2)\notag \\
&=e^{-i(\varphi      -\alpha p)\tilde N}e^{ip \hat Q(q-\alpha n)} 
e^{in\hat \Theta(q -\alpha n)} |\phi(q-\alpha n)\rangle +O(pn, n^2).\label{eq:phi'} 
\end{align}
If $[\tilde N,\hat Q]\neq 0$, terms of $O(p^2)$ remain in Eq.(\ref{eq:phi'}).

The above transformation (\ref{eq:hin_gauge_q})-(\ref{eq:hin_gauge_phi}) is a linear transformation.
In this section, we first consider some linear transformations among $(q^i, p_i)$.
$G$ giving such transformations is a quadratic of $(q^i, p_i)$.
Let $\alpha$ be a constant that is possibly  dependent on time 
but independent of $(q^i, p_i)$ hereinafter.
In the subsections \ref{sec:Ex1}-\ref{sec:Ex4},
we  consider some examples of simple linear gauge transformations.
These examples are useful to understand the possible form of the
general gauge transformation, which is given in the subsection \ref{sec:general}.

\subsection{Example 1 : $G = \epsilon n= \alpha p n$ } \label{sec:Ex1}

This $G$ gives the transformation
\begin{align}
\delta q &= \left\{q,  \alpha p n\right\} = \alpha n, \,\,\, \delta p=0, \\
\delta \varphi &=\left\{\varphi,  \alpha p n\right\} = \alpha p, \,\,\, \delta n=0.
\end{align}
As we  shall see below, it coincides with
the transformation in Ref. \cite{Hinohara2007} including the sign of $\alpha$.
As discussed in Ref. \cite{Hinohara2007}, the basic equations of the ASCC method are invariant 
under 
\begin{align}
 \lambda(q) & \rightarrow  \lambda(q) +\alpha \partial_q V(q), \label{eq:hin3.5c}\\
\partial_q \lambda(q) & \rightarrow  \partial_q \lambda(q) +\alpha C(q). \label{eq:hin3.5d}
\end{align}
if
\begin{equation}
[\hat Q, \tilde N]=0 .
\end{equation}
Although it is not emphasized in Ref. \cite{Hinohara2007},
it is noteworthy that 
\begin{align}
 C(q)=\partial_q^2V+\frac{1}{2B(q)}\partial_qB \partial_q V
\end{align}
is the covariant derivative of $\partial_q V(q)$,
so the transformation of $\partial_q \lambda$
is consistent with that of $\lambda$.

We shall ascertain whether the equation of collective submanifold is
gauge invariant under this transformation.
As mentioned above, the gauge symmetry arises from $n=0$ and
$\dot n=0$.
Therefore, to see the gauge invariance of the equation of collective submanifold,
we drop the term of $\frac{\partial \mathcal{H}}{\partial \varphi}=\dot n=0$,
and set $\frac{\partial \mathcal{H}}{\partial n}=\lambda$ in Eq. (\ref{eq:coll.man2}) 
%with %has the  gauge symmetry. 
It should be noted that, 
%the operators are transformed as 
in correspondence with
\begin{align}
q^\prime&=q+\alpha n,   \label{eq:q' ex1}\\ 
\varphi^\prime&=\varphi+\alpha p \label{eq:phi' ex1},
\end{align}
the differential operators are transformed as
\begin{align}
\partial_{p^\prime}&=\partial_p-\alpha \partial_\varphi, \\ 
\partial_{n^\prime}&=\partial_n-\alpha \partial_q .
\end{align}
%in association  with
This transformation of the differential operators implies 
\begin{align}
\mathring{Q} \rightarrow \mathring{Q} +\alpha \tilde{N}, \label{eq:Q' ex1}\\
\mathring{\Theta} \rightarrow \mathring{\Theta} +\alpha \mathring{P} \label{eq:Theta' ex1}.
\end{align}
In association with this transformation, 
$\hat Q(q)$ and $\hat \Theta(q)$ are transformed as
\begin{align}
\hat{Q} \rightarrow \hat{Q} +\alpha \tilde{N}, \label{eq:^Q' ex1}\\
\hat{\Theta} \rightarrow \hat{\Theta} +\alpha \hat{P}. \label{eq:^Theta' ex1}
\end{align}
This is exactly the transformation found in Ref. \cite{Hinohara2007}.

Then, Eq. (\ref{eq:coll.man2}) is transformed as below:
\begin{align}
& \delta \langle \phi(q,p,n)|\hat H -
  \frac{\partial \mathcal{H}}{\partial p}\mathring{P}
- \frac{\partial \mathcal{H}}{\partial q}\mathring{Q}
- \frac{\partial \mathcal{H}}{\partial n}\mathring{N}
|\phi(q,p,n) \rangle  \notag \\
=&\delta \langle \phi(q,p,n)|\hat H -
  B(q)p\mathring{P}
- \left(\frac{1}{2}\partial_q B(q) p^2+\partial_q V(q) +\partial_q\lambda(q)n \right)\mathring{Q}
- \lambda(q)\mathring{N}
|\phi(q,p,n) \rangle  \notag \\
%\rightarrow 
%&\delta \langle \phi(q+\alpha n,p,n)|\hat H - B(q+\alpha n)p\mathring{P} \notag \\
%&- \left(\frac{1}{2}\partial_q B(q+\alpha n) p^2+\partial_q V(q+\alpha
% n)+\partial_q\lambda(q+\alpha n)n \right)(\mathring{Q}+\alpha
% \mathring{N}) \notag \\
%&- \lambda(q+\alpha n)\mathring{N}|\phi(q+\alpha n, p,n) \rangle  \notag \\
\rightarrow
&\delta \langle \phi(q^\prime, p,n)|\hat H - B(q^\prime)p\mathring{P} %\notag \\
- \left(\frac{1}{2}\partial_q B(q^\prime) p^2+\partial_q
 V(q^\prime)+\partial_q\lambda(q^\prime)n \right) \mathring{Q} \notag \\
&- (\lambda(q^\prime)+\alpha \partial_q V(q^\prime)
 +\frac{1}{2}\alpha\partial_q B(q^\prime) p^2+\alpha\partial_q \lambda(q^\prime)n) \mathring{N}|\phi(q^\prime , p,n) \rangle  \notag 
\end{align}
It is  seen that the Lagrange multiplier is replaced with
\begin{align}
\lambda(q) &\rightarrow \lambda (q^\prime)+\alpha \partial_q V(q^\prime) +\frac{1}{2}\alpha\partial_q B(q^\prime) p^2+\alpha\partial_q\lambda(q^\prime)n\notag\\
&\approx \lambda (q)+\alpha \partial_q V(q)  +\frac{1}{2}\alpha\partial_q B(q) p^2.
\end{align}
This implies that the equation of collective submanifold is
invariant under this gauge transformation with the replacement of the
Lagrange multiplier
\begin{align}
\lambda(q) \rightarrow \lambda (q)-\alpha \partial_q V(q) -\frac{1}{2}\alpha\partial_q B p^2 \label{eq:lambda' ex1}.
\end{align}
This transformation (\ref{eq:lambda' ex1}) is also obtained 
by applying  the transformation (\ref{eq:^Theta' ex1}) 
to the definition of the Lagrange multiplier $\lambda$ (\ref{eq:lambda def}).
From the observation that the operator transformation (\ref{eq:^Q' ex1}) - (\ref{eq:^Theta'
ex1}) is equivalent to the transformation 
of the arguments of the state vector (\ref{eq:hin_gauge_q})- (\ref{eq:hin_gauge_phi}),
one can see that the effect of the gauge transformation (\ref{eq:q'
ex1})-(\ref{eq:phi' ex1}) 
and that of the resulting operator transformation (\ref{eq:^Q' ex1})-(\ref{eq:^Theta' ex1})
cancel out with each other, which leads to the gauge invariance.
(This can be understood most clearly in Example 4 below.) 

We shall focus on the point that $\lambda(q)$ is also transformed 
in association with the gauge transformation.
It is a matter of course because the degrees of freedom of the gauge
transformation are those of the Lagrange multipliers for the first-class
primary constraints~\cite{Sugano1991}.
It should be noted that Eq. (\ref{eq:lambda' ex1}) contains a term of $O(p^2)$.
In the equation of collective submanifold, the gauge symmetry is kept  
by including the change in $\lambda$ up to $O(p^2)$ as we have seen above.
After the adiabatic expansion, however,
it is not settled with only by the change of $\lambda$.
While, in the moving-frame HFB equation (\ref{eq:moving-frame HFB}),
the effect of the transformation of $\hat Q$ is canceled out 
with the transformation of $\lambda$,
in the moving-frame QRPA equation of $O(p^2)$ (\ref{eq:moving-frame QRPA1}),
the term of $\tilde N$ comes up from the gauge transformation of
$\hat Q$, which was supposed to be canceled by replacing $\lambda$.
To cancel this $\tilde N$ term, 
we need to change $\partial_q \lambda$ also. 
(Nevertheless, it is natural that $\partial_q \lambda$ is transformed 
in association with the transformation of $\lambda$.)
We shall focus on the equation of $O(p^2)$ 
in the adiabatic expansion of the equation of collective submanifold
((2.36) in Ref. \cite{Matsuo2000}) in turn.
The moving-frame QRPA (\ref{eq:moving-frame QRPA2}) is derived from this equation
\begin{align}
 \delta\langle \phi(q)|\frac{1}{2}[[\hat H-\lambda\tilde N, \hat
 Q],\hat Q]-B\partial_q\hat Q+\frac{1}{2}\partial_qB\hat
 Q|\phi(q)\rangle=0.\label{eq:M2.36}
\end{align}
This equation does not satisfy the gauge symmetry,
because there is no $\tilde N$ term, which can cancel 
the transformation $\hat Q \rightarrow \hat Q+\alpha \tilde N$.
As the $\tilde N$ term is brought in the moving-frame QRPA equation 
by rewriting Eq. (\ref{eq:M2.36})  using the moving-frame HFB equation, 
the gauge symmetry is satisfied with the replacement of
$\partial_q\lambda$.
In the sense that an equation derived from a more fundamental equation
should keep the symmetry of the original equation as much as possible,
the moving-frame QRPA equation (\ref{eq:moving-frame QRPA2}) 
is better than Eq. (\ref{eq:M2.36})
As seen also in a later example, 
the gauge symmetry is (partially) broken by the adiabatic expansion. 
%In other words, the adiabatic expansion (parially) fixes the gauge.
This gauge symmetry holds irrespectively of the order of $n$ since
$n=0$.
Thus it has been shown that the symmetry under the transformation
(\ref{eq:hin3.5a}) and (\ref{eq:hin3.5b})
found in Ref. \cite{Hinohara2007} is a gauge symmetry.

\subsection{Example 2 : $G = \epsilon n= \alpha n^2/2$ }
This $G$ generates the transformation
\begin{align}
\delta q &= 0, \,\,\,  \delta  p=0, \\
\delta \varphi &=\left\{\varphi,\alpha  n^2/2\right\} =  \alpha n, \,\,\,  \delta  n = \left\{n,  \alpha  n^2/2\right\}=0 .
\end{align}
Corresponding to $\varphi \rightarrow \varphi +\alpha n$, we consider
\begin{align} 
%hat Q      & \rightarrow  \hat Q   +\alpha \tilde N  \\
\hat \Theta & \rightarrow \hat \Theta + \alpha \tilde N \label{eq:Theta' ex2}.
\end{align}
It can be easily verified that 
the canonical-variable conditions, the moving-frame HFB and the moving-frame
QRPA equations are invariant under this transformation.
(The canonicity conditions always are met
for $(q^{\prime i},p^\prime_i)$ which are canonically transformed from 
the canonical variables $(q^i,p_i)$ which satisfy the canonicity conditions
\cite{Yamamura1987}.
This applies in the following examples also.)
%and it can be seen from that 
One can see that this transformation does not affect the Lagrange multiplier $\lambda$.
When the transformation (\ref{eq:Theta' ex2}) is applied to 
(\ref{eq:lambda def}), $\lambda$ remains unchanged. 
In Examples 1, 2, and 4,
$\lambda$  is transformed in association with
the transformation of 
$\varphi$ and one of $(q,p,n)$.
%are transformed and 
%it associates with the change of 
As we are considering the potential $V=V(q)$,
in the Lagrange formalism, our system is a one-dimensional one
with the degrees of freedom $(q, \dot q)$ only.
The reason why $\lambda$ does not change under the gauge transformation 
in this example has something to do with 
the fact that we have regarded our one-dimensional system 
as a two-dimensional one with $V=V(q,\varphi)$ virtually
and introduced the constraint $p_2=n=0 $ in the Hamilton formalism.

To clarify this point, 
we shall consider a system whose potential depends on $\varphi$
and denote the potential as $\tilde V$ instead of $V$.
From Eq. (\ref{eq:n dot}),
for the primary constraint $\Phi^1=n=0$ to be always met, 
$\partial_\varphi \tilde V\approx 0$ is required.
Here, we shall consider the potential below:
\begin{align}
 \tilde V = \eta \varphi   + V(q),
\end{align}
with an infinitesimally small constant $\eta$.
$\tilde V$ reduces to $V(q)$ in the limit of $\eta \rightarrow 0$:
\begin{align}
\lim _{\eta \rightarrow 0} \tilde V =  V(q).   
\end{align}
As $\partial_\varphi \tilde V=\eta$, 
the consistency condition for $n=0$ is satisfied 
in the limit of $\eta \rightarrow 0$.
In this limit, the Lagrangian is reduced to
\begin{align}
 L=\frac{1}{2B(q)} \dot q ^2 -\tilde V(q,\varphi) \rightarrow
 \frac{1}{2B(q)} \dot q ^2-V(q) \,\,\, (\eta \rightarrow 0).
\end{align}
The Hamiltonian is given by
\begin{align}
 H=\dot q^i p_i -L =\frac{1}{2}B(q)p_1^2+\tilde V
 =\frac{1}{2}B(q)p_1^2+\eta \varphi +V(q),
\end{align}
and the total Hamiltonian is
\begin{align}
 H_T=\frac{1}{2}B(q)p_1^2+\eta \varphi +V(q) +\lambda n.
\end{align}
The generator of the gauge transformation $G$ fulfills the stationary
condition as
\begin{align}
& \{G,n\}=\partial_\varphi\epsilon n \approx 0,\\
& \{G, H \}+\partial_t G =   \epsilon \{n, H\}+ \{\epsilon , H\}n
 + \partial_t \epsilon n \approx \epsilon \eta  n
 \approx 0.
\end{align}
(We have assumed that $\epsilon$ is a regular function of $n$.)
$G=\alpha n^2/2$ generates the transformation
$\delta \varphi =\{ \varphi, \alpha n^2/2 \}=\alpha n$,
and then the potential is transformed as
\begin{align}
\tilde V\rightarrow &\tilde V(q,\varphi+\delta \varphi) \notag \\
=& V(q) + \eta \varphi + \eta \delta \varphi   \notag \\
=& V(q) + \eta \varphi 
 +  \alpha \eta n  .
\end{align}
One can see that, with a replacement of  $\lambda$ 
\begin{align}
\lambda \rightarrow & \lambda -  \eta \alpha,   
\end{align}
the basic equations of the ASCC method are invariant. 
Because we are now considering the $\eta \rightarrow 0$ limit in this example,
$\lambda$ is not affected by the gauge transformation.

\subsection{Example 3: $G =  \epsilon n= \alpha qn$}

This $G$ generates the transformation
\begin{align}
\delta q &= 0, \,\,\,  \delta  p=-\alpha n, \\
\delta \varphi &=\alpha q, \,\,\,  \delta  n =0. 
\end{align}
%$p \rightarrow p -\alpha n$ and $\varphi \rightarrow \varphi +\alpha q$
 Corresponding to this transformaion, we consider
\begin{align} 
\hat P      & \rightarrow  \hat P   -\alpha \tilde N, \label{eq:P' ex3}\\
\hat \Theta & \rightarrow \hat \Theta + \alpha \hat Q.
\end{align}
The moving-frame HFB equation (\ref{eq:moving-frame HFB}) 
and the canonical-variable conditions  are
invariant under this transformation.
If we assume that $[\tilde N,\hat Q]=0$ as in Example 1,
the moving-frame QRPA equation of $O(p^2)$ (\ref{eq:moving-frame QRPA2}) is invariant but
the moving-frame QRPA equation of $O(p)$ (\ref{eq:moving-frame QRPA1})
is not; it is tranformed as
\begin{equation}
 \delta \langle \phi(q)|[\hat H-\lambda \tilde N -\partial_q V\hat Q, i\hat Q] -B(q)\hat P
  +\alpha B(q)\tilde N 
|\phi(q)\rangle =0. \label{eq:mfQRPA1'Ex3}
\end{equation}
As we shall see, the equation of collective submanifold is invariant under
\begin{align} 
\mathring P      & \rightarrow  \mathring P   -\alpha \tilde N,  \\
\mathring \Theta & \rightarrow \mathring \Theta + \alpha \mathring Q.
\end{align}

Note that only the A-part of $\tilde N$ contributes to the
above equation (\ref{eq:mfQRPA1'Ex3}). 
(We call the $a^\dagger a^\dagger$ and $aa$ part of an operator
the A-part and the $a^\dagger a$ part the B-part.)
Therefore, we instead consider 
\begin{align} 
\hat P      & \rightarrow  \hat P  -\alpha \tilde N^B, \label{eq:P'B ex3} \\ 
\hat \Theta & \rightarrow \hat \Theta + \alpha \hat Q \label{eq:Theta'  ex3}. 
\end{align}
Here, $\tilde N^B$ denotes the B-part of $\tilde N$.
The moving-frame HFB equation (\ref{eq:moving-frame HFB}), 
the moving-frame QRPA (\ref{eq:moving-frame QRPA1})
are invariant under the transformation above.
The moving-frame QRPA equation (\ref{eq:moving-frame QRPA2}) reads
\begin{align}
 \delta \langle \phi(q)|
[\hat H -\lambda \tilde N -\partial_q V \hat Q, \hat P -\alpha \tilde N^B] + 
\cdots |\phi(q)\rangle =0.
\end{align}
Because the A-part of $\hat H_M=\hat H -\lambda \tilde N -\partial_q V
\hat Q$ 
vanishes from the moving-frame HFB equation,
$[\hat H_M, \tilde N^B]$ does not contribute.
After all, the moving-frame HFB \& QRPA equations are invariant 
even if $[\tilde N,\hat Q] \neq 0$.

However, among the canonical-variable conditions, 
only (\ref{eq:O(n) can.var.cond 1})
is not invariant as shown below:
\begin{align}
&\langle \phi(q)|\,[\hat P, \hat \Theta] \,|\phi(q)\rangle =0\notag \\
\rightarrow
&\langle \phi(q)|\,[\hat P, \hat \Theta] \,|\phi(q)\rangle
-\alpha \langle \phi(q)|\,[\tilde N^B, \hat \Theta] \,|\phi(q)\rangle \notag\\
&\,\,\,\,\,\,\,\,\,+\alpha \langle \phi(q)|\,[\hat P, \hat Q] \,|\phi(q)\rangle
+\alpha^2 \langle \phi(q)|\,[\tilde N^B, \hat Q] \,|\phi(q)\rangle
= -\alpha i.
\end{align}
We have used that the fourth term vanishes because $\tilde N^B$ and $\hat Q$
are time-even.
The other canonical-variable conditions are invariant.

To sum up, under the transformation (\ref{eq:P'B ex3})-(\ref{eq:Theta'  ex3}),
the ASCC basic equations are not fully invariant, 
but the moving-frame HFB \& QRPA equations are invariant.
Therefore, if we allow $\hat P$ to have a B-part,
we need to take into account this gauge degree of freedom in
solving the moving-frame HFB \& QRPA equations.

We shall see this gauge transformation at the level of the equation of
collective submanifold.
At the level of the equation of
collective submanifold (\ref{eq:coll.man2}), the transformation reads
\begin{align}
 -Bp(\mathring{P} -\alpha \tilde N)-\lambda\tilde N =
 -Bp\mathring{P} -(\lambda-\alpha Bp )\tilde N,   
\end{align}
and 
\begin{align}
\lambda  \rightarrow \lambda+\alpha Bp \label{eq:lambda' ex3}
\end{align}
makes the equation of collective submanifold invariant.
Eq. (\ref{eq:lambda' ex3}) can be also obtained by applying (\ref{eq:Theta'
ex3}) to Eq. (\ref{eq:lambda def}).
This is the reason why the moving-frame QRPA equation
(\ref{eq:moving-frame QRPA1}) is not invariant 
under the transformation (\ref{eq:P' ex3}).
More precisely,
while in the equation of collective submanifold the change of $\hat P$
is absorbed by (\ref{eq:lambda' ex3}),
in the moving-frame HFB \& QRPA equations,
this change of the Lagrange multiplier $\lambda$ is not included
because the equation of collective submanifold is divided 
into the three equations depending on the order of $p$.
As seen in this example, the gauge symmetry is partially broken by the
adiabatic expansion.
In other words, the adiabatic expansion partially fixes the gauge.

\subsection{Example 4 : $G = \epsilon n = \alpha \varphi n$} \label{sec:Ex4}

This $G$ generates 
\begin{align}
\delta q &= 0, \,\,\,  \delta  p= 0, \\
\delta \varphi &=\alpha \varphi , \,\,\,  \delta  n =-\alpha n. 
\end{align}
This is the following point transformation.
\begin{align}
\varphi &\rightarrow \varphi +\alpha \varphi = e^\alpha \varphi, \\
n &\rightarrow n -\alpha n = e^{-\alpha} n
\end{align}
The corresponding transformation of the operators is given by
\begin{align} 
\hat \Theta & \rightarrow  \hat \Theta + \alpha \hat \Theta =e^{\alpha}\hat \Theta, \label{eq:Theta'  ex4}\\
\tilde N    & \rightarrow  \tilde N   -\alpha \tilde N = e^{ -\alpha}\tilde N. 
\end{align}
As the state vector is given by
\begin{align}
|\phi(q,p,\varphi, n) \rangle =e^{-i \varphi \tilde N}e^{i(p\hat Q +n\hat\Theta)} |\phi(q) \rangle,  
\end{align}
one can easily see that the scale transformation of 
$(\varphi,n)$ and that of $(\hat \Theta,\tilde N)$ cancel with each other.

The canonical-variable conditions are invariant under this gauge transformation.
The moving-frame HFB \& QRPA equations are also invariant with
\begin{align}
\lambda(q) &\rightarrow \lambda(q)e^{\alpha} \label{eq:lambda' ex4}\\ 
\partial_q \lambda(q) &\rightarrow \partial_q \lambda(q)e^{\alpha} .
\end{align}
The transformation (\ref{eq:lambda' ex4}) can be also obtained by applying (\ref{eq:Theta'
ex4}) to (\ref{eq:lambda def}).

It is worth noting the following remarks here:
%(1) In Examples 1, 2, and 3, $alpha$ can be finite, while, in Example 4,
%    for finite $\alpha$, it should be in the form of $e^{\pm \alpha}$ as above.
(i) Because $\tilde N$ is a known operator, this gauge degree of freedom
does not cause numerical instability in actual calculation of the moving-frame
HFB \& QRPA equations.
(ii) The generator of the similar point transformation for $\hat Q,\, \hat P$
is given by $G=\alpha qp$. However, it does not satisfy the stationary
condition, and thus it is not a gauge transformation.
The point transformation of $(q, p)$ is just a matter of the choice of
coordinate system, and it does not mean non-uniqueness of the solution to
the equation of motion. 

We show the generating functions of the canonical transformations for 
the time-independent gauge transformations in Examples 1-4.
\begin{align}
 \text{Example 1 : } S&= -p_i \{q^i, -\alpha pn\}-\alpha pn=\alpha pn =-G. \\
 \text{Example 2 : } S&= -p_i \{q^i, \alpha n^2/2\}+\alpha n^2/2=-\alpha n^2/2 =-G. \\
 \text{Example 3 : } S&= -p_i \{q^i, \alpha qn\}+\alpha qn=-\alpha qn +\alpha qn =0. \\
 \text{Example 4 : } S&= -p_i \{q^i, \alpha \varphi n\}+\alpha \varphi n=-\alpha \varphi n +\alpha \varphi n = 0.
\end{align}
The gauge transformations in Examples 3 and 4 are point transformations,
so the generating functions are constant as discussed in Ref. \cite{Yamamura1987}.

\subsection{General gauge transformation} \label{sec:general}
In this subsection, we present the general infinitesimal gauge transformation
\begin{align}
 G=\epsilon(q,p,\varphi, n,t)n.
%= \epsilon^\prime(q,p,\varphi, n,t)\delta s
\end{align}
%Under the general gauge transformation, 
We have seen in Sec. \ref{sec:generators} that the coordinates and momenta are transformed as follows:
\begin{align}
 q^\prime      &= q + n\partial_p \epsilon, \\
 p^\prime      &= p - n\partial_q \epsilon, \\
 \varphi^\prime &= \varphi + \epsilon + n\partial_n\epsilon, \\
 n^\prime      &= n ( 1 -\partial_\varphi \epsilon)
\end{align}
We denote $(q,\varphi,p,n)$ as $\xi^i$ collectively, and then
the Jacobi matrix for this transformation is 
written as
\begin{align}
%\frac{\partial( q^\prime, \varphi^\prime, p^\prime, n^\prime)}{\partial
% (q,\varphi, p, n)}=
\left(\frac{\partial \xi^{\prime i} }{\partial \xi^j}\right)
&=I+
 \begin{pmatrix}
n\partial_q\partial_p\epsilon  &&    n\partial_\varphi\partial_p\epsilon
  &&    n\partial_p^2\epsilon  &&    \partial_p\epsilon+n\partial_n\partial_p\epsilon  \\    
\partial_q\epsilon +n\partial_q\partial_n\epsilon  && \partial_\varphi+n\partial_\varphi\partial_n\epsilon  &&    
\partial_p \epsilon+n\partial_p\partial_n\epsilon  &&    2\partial_n\epsilon+n\partial_n^2\epsilon  \\    
-n\partial_q^2\epsilon  &&    -n\partial_\varphi\partial_q\epsilon
  &&   - n\partial_p\partial_q\epsilon  &&    -\partial_q\epsilon-n\partial_n\partial_q\epsilon  \\    
-n\partial_q\partial_\varphi\epsilon  &&    -n\partial_\varphi^2\epsilon
  &&   - n\partial_p\partial_\varphi\epsilon  &&    -\partial_\varphi\epsilon-n\partial_n\partial_\varphi\epsilon    
 \end{pmatrix}\notag \\
&\approx I+
\begin{pmatrix}
0                    &&    0                       &&           0             &&    \partial_p\epsilon  \\    
\partial_q\epsilon   && \partial_\varphi\epsilon   &&    \partial_p \epsilon  &&    2\partial_n\epsilon \\    
0                    &&    0                       &&           0             &&    -\partial_q\epsilon  \\    
0                    &&    0                       &&           0             &&    -\partial_\varphi\epsilon    
\end{pmatrix}\notag \\
&:=I+M,
\end{align}
where $I$ is the unit matrix. By omitting the higher-order terms, we obtain
\begin{align}
\partial_{\xi^{j\prime}}=\frac{\partial \xi^i}{\partial \xi^{j \prime}}\frac{\partial }{\partial \xi^i} 
=^t(I+M)^{-1}\frac{\partial}{\partial \xi^{i}}=^t(I-M)\frac{\partial}{\partial \xi^{i}},
\end{align}
i. e., 
\begin{align}
\begin{pmatrix}
 \partial_{q^\prime} \\
 \partial_{\varphi^\prime} \\
 \partial_{p^\prime} \\
 \partial_{n^\prime} \\
\end{pmatrix}
=
\begin{pmatrix}
 \partial_{q} \\
 \partial_{\varphi} \\
 \partial_{p} \\
 \partial_{n} \\
\end{pmatrix}
+
\begin{pmatrix}
0                    && -\partial_q\epsilon        &&    0              &&           0       \\  
0                    && -\partial_\varphi\epsilon  &&    0              &&           0       \\       
0                    &&   -\partial_p \epsilon     &&    0              &&           0        \\              
 -\partial_p\epsilon  &&    -2\partial_n\epsilon &&       \partial_q\epsilon  &&   \partial_\varphi\epsilon 
\end{pmatrix}
\begin{pmatrix}
 \partial_{q} \\
 \partial_{\varphi} \\
 \partial_{p} \\
 \partial_{n} \\
\end{pmatrix},
\end{align}
from which we read 
\begin{align}
\mathring{P}  & \rightarrow \mathring {P} -\partial_q\epsilon \tilde N, \\
\tilde N      & \rightarrow  (1 - \partial_\varphi \epsilon ) \tilde N, \\
\mathring{Q}  & \rightarrow \mathring {Q} + \partial_p\epsilon \tilde N, \\
\mathring{\Theta} & \rightarrow (1+\partial_\varphi
 \epsilon)\mathring{\Theta} + \partial_p\epsilon \mathring{P}
 +2\partial_n\epsilon \tilde N 
+ \partial_q\epsilon \mathring{Q}. 
\end{align}
Under this transformation, the equation of collective submanifold (\ref{eq:coll.man2}) is
transformed as
\begin{align}
&\delta \langle \phi(q,p,n)|\hat H -
  \frac{\partial \mathcal{H}}{\partial p}\mathring{P}
- \frac{\partial \mathcal{H}}{\partial q}\mathring{Q}
- \frac{\partial \mathcal{H}}{\partial N}\mathring{N}
|\phi(q,p,N) \rangle  \notag  \\
%\rightarrow & 
%\delta \langle \phi(q^\prime,p^\prime ,n^\prime)|\hat H -
%  \frac{\partial \mathcal{H^\prime}}{\partial p}(\mathring {P} -\partial_q\epsilon \tilde N )
%- \frac{\partial \mathcal{H^\prime}}{\partial q}(\mathring {Q} +
% \partial_p\epsilon \tilde N ) \notag \\
%&- \frac{\partial \mathcal{H^\prime}}{\partial N}(1 - \partial_\varphi \epsilon ) \tilde N 
%|\phi(q^\prime,p^\prime,n\prime) \rangle    \notag \\
\rightarrow & 
\delta \langle \phi(q^\prime,p^\prime ,n^\prime)|\hat H 
- B(q^\prime)p^\prime(\mathring {P} -\partial_q\epsilon \tilde N )
- (\frac{1}{2}\partial_{q^\prime} B(q^\prime)p^{\prime 2}+\partial_{q^\prime} V(q^\prime)+\partial_{q^\prime}\lambda(q^\prime)n^\prime )(\mathring {Q} +
 \partial_p\epsilon \tilde N ) \notag \\
&- \lambda(q^\prime)(1 - \partial_\varphi \epsilon ) \tilde N 
|\phi(q^\prime,p^\prime,n\prime) \rangle  \notag \\  
\approx & 
\delta \langle \phi(q, p, n)|\hat H 
- B(q)p(\mathring {P} -\partial_q\epsilon \tilde N )
-  (\frac{1}{2}\partial_{q} B(q)p^{2}+\partial_{q} V(q))(\mathring {Q} +
 \partial_p\epsilon \tilde N ) \notag \\
&- \lambda(q)(1 - \partial_\varphi \epsilon ) \tilde N 
|\phi(q, p, n) \rangle \notag \\
= & 
\delta \langle \phi(q, p, n)|\hat H 
- B(q)p\mathring {P} 
-  (\frac{1}{2}\partial_{q} B(q)p^{2}+\partial_{q} V(q))\mathring {Q}  \notag \\
&- \left[\lambda(q)(1 - \partial_\varphi \epsilon ) -B(q)p\partial_q\epsilon 
 +(\frac{1}{2}\partial_{q} B(q)p^{2}+\partial_{q} V(q)) \partial_p\epsilon\right]
\tilde N 
|\phi(q, p, n) \rangle .
\end{align}
Thus we can see that the equation of collective submanifold is invariant 
under the general gauge transformation if we replace the Lagrange multiplier 
as
\begin{align}
\lambda(q)\rightarrow \lambda(q) + \partial_\varphi \epsilon \lambda(q) +B(q)p\partial_q\epsilon 
 -(\frac{1}{2}\partial_{q} B(q)p^{2}+\partial_{q} V(q)) \partial_p\epsilon.
\end{align}

As $\tilde N$ is a known operator,
we shall concentrate on the case with $\partial_\varphi\epsilon=0$. 
In that case, under the transformation 
\begin{align}
\hat{P}  & \rightarrow \hat {P} -\partial_q\epsilon \tilde N^B,  \label{eq:P' gen}\\
%\tilde N      & \rightarrow  (1 - \partial_\varphi \epsilon ) \tilde N \\
\hat{Q}  & \rightarrow \hat {Q} + \partial_p\epsilon \tilde N, \\
\hat{\Theta} & \rightarrow \hat{\Theta} + \partial_p\epsilon \hat{P}+2\partial_n\epsilon \tilde N 
+ \partial_q\epsilon \hat{Q} \label{eq:Theta' gen},
\end{align}
the moving-frame HFB \& QRPA equations are invariant if we transform
$\lambda$ and $\partial_q \lambda$ as
\begin{align}
\lambda(q) &\rightarrow \lambda(q) -\partial_{q} V(q)
 \partial_p\epsilon  \label{eq:lambda' gen}, \\
\partial_q \lambda(q) &\rightarrow \partial_q\lambda(q) - C(q)
 \partial_p\epsilon. \label{eq:dqlambda' gen}
\end{align}
Following the discussion in Example 3, we have replaced $\tilde N$ with
$\tilde N^B$ in (\ref{eq:P' gen}).
If $\partial_p\epsilon\neq 0$, $[\tilde N, \hat Q]=0$ is required for
the invariance.
Moreover, if $\partial_q\epsilon\neq 0$, the invariance of the
canonical-variable condition (\ref{eq:O(n) can.var.cond 1}) is broken
as discussed in Example 3.
%Note that, in 
In Example 3, we have considered 
the gauge transformation which mixes 
the time-odd operator $\hat P$ with the time-even operator $\tilde N$
($\alpha$ is assumed to be real).
%$\hat P$ is time-odd, $\tilde N$ is time-even, and $\alpha$ is a
%time-even constant 
If we require that time-odd and time-even operators do not mix,
we can exclude the gauge transformation of Example 3.
However, for the general gauge transformation, 
%(\ref{eq:P' gen})-(\ref{eq:Theta' gen}),
if $\partial_q \epsilon$ is time-odd,
we need to take into account transformations 
which mix $\hat P$ with $\tilde N$ also.

Before ending this section, we present one special case, namely,
the general linear gauge transformation
\begin{align}
q &\rightarrow q + \beta n, \\
p &\rightarrow p - \alpha n, \\
\varphi &\rightarrow \varphi+\alpha q+\beta p +\eta n,
\end{align}
with constant $\alpha, \beta$ and $\eta$.
(From discussion in Example 2, the $\eta$ can be set to $0$ without loss of generality) 
We have ignored
the scale transformation of $\tilde N$ following the discussion in Example 4.
Eqs. (\ref{eq:P' gen})-(\ref{eq:Theta' gen}) and (\ref{eq:lambda' gen})-(\ref{eq:dqlambda' gen}) read 
\begin{align} 
\hat Q    & \rightarrow  \hat Q +\beta \tilde N, \\
\hat P    & \rightarrow  \hat P -\alpha \tilde N^B,  \\
\hat \Theta & \rightarrow \hat \Theta +\alpha \hat Q+\beta \hat P +\eta \tilde N,  
\end{align}
and \begin{align}
 \lambda(q) \rightarrow \lambda(q) - \beta \partial_q V(q), \\
 \partial_q \lambda(q) \rightarrow \partial_q\lambda(q) - \beta C(q), 
 \end{align}
respectively.

\section{Expansion of the collective Hamiltonian up to $O(n^2)$}
\label{sec:O(n^2)}

As we have seen above, the linear term of $n$ in the collective
Hamiltonian plays a key role for the gauge symmetry in the ASCC method. 
In this section, we consider the case where the collective 
Hamiltonian is expanded up to $O(n^2)$.

We shall consider the system whose Lagrangian is given by
\begin{equation}
 L=\frac{1}{2B(q^1)}(\dot q^1)^2+\frac{1}{2D(q^1)}(\dot
  q^2-\lambda(q^1))^2-V(q^1)  \label{eq:L O(n^2)}.
\end{equation}
The rank of the Hessian ($\partial^2L / \partial \dot q^i \partial\dot q^j$) is 2. 
Since the Hessian is not degenerate, there is no
constraint. Hence there is no gauge degree of freedom in this system.
The momenta are calculated as 
\begin{align}
 p_1&=\frac{1}{B(q)}\dot q^1, \\
 p_2&=\frac{1}{D(q)} (\dot q^2 -\lambda(q)),
\end{align}
and then the  Hamiltonian is given by
\begin{align}
 H  &=\dot q^i p_i -L \notag \\
    &= \frac{1}{2} Bp_1^2 +\frac{1}{2}Dp_2^2 +\lambda p_2+V(q^1).
\end{align}
Noting that $q^1=q$ and  $q^2=\varphi$, this is the collective Hamiltonian
expanded up to $O(n^2)$ in the ASCC method.
Therefore, in the case where we take up to the $O(n^2)$ terms in the
ASCC method, there is no gauge symmetry.

The particle number is conserved also in this case.
The Euler-Lagrange equation for (\ref{eq:L O(n^2)}) leads to
\begin{align}
  \frac{d}{dt} \left(\frac{\partial L}{\partial \dot
 q^2}\right)=\frac{dp_2}{dt}=\frac{\partial L}{\partial q^2}=0.
\end{align}
This is the conservation of the momentum conjugate to a cyclic variable, 
and it is a consequence of Noether's first theorem 
for the translation of $q^2=\varphi$.

As long as we consider only the moving-frame HFB \& QRPA equations of 
$O(1), O(p^1)$ and $O(p^2)$ (\ref{eq:moving-frame
HFB})-(\ref{eq:moving-frame QRPA2}), a gauge symmetry appears even if we
expand the collective Hamiltonian up to $O(n^2)$.
This is because it is equivalent to setting $n=0$ to ignore the $O(n)$ and higher-order expansions 
in the equation of collective submanifold.
As discussed above, when
\begin{equation}
 \frac{\partial \mathcal{H}}{\partial n}=\lambda(q),
\end{equation}
the equation of collective submanifold possesses a gauge symmetry.
Even if we expand $\mathcal{H}$ up to $O(n^2)$, its derivative 
\begin{equation}
 \frac{\partial \mathcal{H}}{\partial n}=\lambda(q)+D(q)n
\end{equation}
coincides with 
$\frac{\partial \mathcal{H}}{\partial n}=\lambda(q)$
as long as we ignore the linear and higher-order terms of $n$.  
Therefore, the gauge symmetry appears in the equations of $O(1),O(p^1)$ and $O(p^2)$.

This can be understood also in the following way.
As $n=const.$, we can choose an inertial frame in which $n=0$.
Because the $\dot q^2$ term vanishes in  (\ref{eq:L O(n^2)}),
this case reduces to the case of the singular Lagrangian we discussed 
in the beginning of this paper, and thus the gauge symmetry occurs. 
Choosing the inertial frame such that $n=0$ 
can be interpreted as a constraint.
%In other words, we impose a constraint that ``we choose the inertial frame
%such that $p_2=0$``, and it gives the gauge symmetry.

\section{Concluding remarks}

In this paper, we have investigated the gauge symmerty in the ASCC method
on the basis of the Dirac-Bergmann theory of constrained systems.
We have seen that the transformation found in  Ref. \cite{Hinohara2007} 
is one example of the gauge transformations and that
the gauge symmetry originates from the constraint $n=0$ 
in the collective Hamiltonian. 
We have discussed the symmetry of the ASCC equations 
under the possible gauge transformations
and the validity of the adiabatic expansion. 
While the equation of collective submanifold is invariant 
under the general gauge transformation, 
the gauge symmerty is partially broken by the adiabatic expansion
at the level of the moving-frame HFB \& QRPA equations.

As we have seen, in either case of the expansion up to $O(n)$
or up to $O(n^2)$, the particle number is conserved.
Eq. (\ref{eq:hin2.3}) plays an important role for the number conservation.
%However, 
The difference between the two cases may be worth noting.
Whereas, in the $O(n)$ case, $n=0$ and $\dot n=0$ are the constraint 
and its consistency condition, respectively,
$\dot n=0$ is the consequence of the Noether's theorem about the
translation of $\varphi$ in the $O(n^2)$ case.

Because we are considering a constrained system with $n=0$,
it is well justified to adopt the 
$O(1), O(p)$ and $O(p^2)$ equations as the basic equations
of the ASCC method.
For applications of the ASCC method to number non-conserving phenomena 
such as pairing vibration,
the expansion up to $O(n^2)$ will be needed.
It will be investigated in a future publication.

For the canonical-variable conditions, we have listed Eqs. 
(\ref{eq:O(n) can.var.cond 1}) and (\ref{eq:O(n) can.var.cond 3}),
but they are actually the equations of $O(n)$.
It may seem more consistent to adopt the canonical-variable conditions of 
only $O(1)$ and $O(p^1)$ ( and $O(p^2)$ ) 
following the orders of the moving-frame HFB \& QRPA equations.
We have adopted them, however, because
the $O(n)$ canonical-variable conditions are necessary
for the discussion on the gauge invariance.
 
We have discussed the gauge symmetry associated with the first-class
constraint $n=0$ in this paper.
It is known \cite{Maskawa1976} that,
when there are $A$ constraints of the first class and $2T$ constraints
of the second class in the $2N$ dimensional phase space,
with an appropriate variable transformation $(q^i,p_i)\rightarrow (Q^i, P_i)$,
we can choose canonical variables such that
\begin{minipage}[c]{\textwidth}
\begin{align}
P_a &\approx 0, \,\,\,\,(a=1, \dots A), \\
Q^t &\approx 0, P_ t  \approx 0, (t=1, \dots T) .
\end{align} 
\end{minipage}
%(The other variables are not constrained.)
In the case of the ASCC method, $A=1$ and $P_1=n$.
In the ASCC method, we assume that the motion of the system is confined 
onto the collective submanifold embedded in the TDHFB phase space. 
Although there appears no second-class constraint 
in the formulation of the ASCC method,
the above assumption itself may correspond to the second-class constraints.
The assumption on the collective submanifold 
may be rephrased as the assumption that
there are an even number of the second-class constraints defined 
in the TDHFB phase space, 
which restrict the motion of the system within the collective submanifold.

In Sec. \ref{sec:HT} we started from the Hamiltonian and then considered
the corresponding Lagrangian, 
which gives (\ref{eq:HT}) as the total Hamiltonian.
Let us consider the structure of the theory 
assuming that we were given the Lagrangian first, although in the
ASCC method the (total) Hamiltonian is first given. 
As discussed in Ref. \cite{Sugano1983}, the system whose Lagrangian is given
by (\ref{eq:Lagrangian}) is strongly reducible to a one-dimensional system
if $\partial_\varphi V=0.$
It is also seen that there appears no secondary constraint 
from the condition for the system to be gauge-invariant.
As $\partial_\varphi V=0$ leads to $V=V(q)$,  the degree of freedom
of $(\varphi, \dot \varphi)$ disappears from the Lagrangian.
%If we presumed to treat the system as a two-dimensional one although 
%it is strongly reducible to the one-dimensional one,
Although we could treat the system as a one-dimensional one, 
we dared to regard it as a two-dimensional one 
and moved to the Hamilton formalism.
The resulting total Hamiltonian (\ref{eq:HT}) is 
of the form of a weakly reducible system. 
It should be noted, however, 
that the degrees of freedom of $(q,p)$ and $(\varphi, n)$ are not completely
decoupled because $\lambda=\lambda(q)$ in the ASCC method.

In this paper, 
we have considered the one-dimensional single-component ASCC method 
for simplicity.
The extension to the multi-dimensional and multi-component 
cases is straightforward.

\section*{Acknowledgements}
The author thanks K.~Matsuyanagi, M.~Matsuo, T.~Nakatsukasa, and N.~Hinohara for
their useful comments.
The author is supported by the Special Postdoctoral Researcher Program
of RIKEN.

%\section*{References}
%\label{sec3.18}

%\bibliography{gaugeASCC.bib}
%\end{document}

\end{document}